\documentclass[prl,twocolumn,showpacs,floatfix,amsmath,amssymb,superscriptaddress]{revtex4}
\usepackage[dvips]{graphicx}
\usepackage{latexsym}
\usepackage{graphicx}
\usepackage{times}
\usepackage{amsmath}
\usepackage{dcolumn}
\usepackage{subfigure}
\usepackage{latexsym,amsmath,amssymb,bm,euscript}

\newcommand{\be}{\begin{equation}}
\newcommand{\ee}{\end{equation}}

\newcommand{\ba}{\begin{eqnarray}}
\newcommand{\ea}{\end{eqnarray}}

\begin{document} 
\title{Theory of
dissipationless Nernst effects.} 
\author{Doron L. Bergman}\email{doron@caltech.edu}
\affiliation{Department of Physics, California Institute of Technology,  Pasadena, CA 91125}
\author{Vadim Oganesyan} \email{oganesyan@mail.csi.cuny.edu}
\affiliation{Department of Engineering Science and Physics, College of Staten Island, CUNY, Staten Island, NY 10314}
\date{\today} 
\begin{abstract}
We develop a theory of transverse thermoelectric (Peltier) conductivity, $\alpha_{xy}$, in finite magnetic field -- this particular conductivity is often the most important contribution to the Nernst thermopower. 
We demonstrate that $\alpha_{xy}$ of a free electron gas can be expressed purely and exactly as the entropy per carrier irrespective of temperature (which agrees with seminal Hall bar result of Girvin and Jonson). 
In two dimensions we prove the universality of this result in the presence of disorder which allows explicit demonstration of a number features of interest to experiments on graphene and other two-dimensional materials.
We also exploit this relationship in the low field regime and to analyze the rich singularity structure in $\alpha_{xy}(B,T)$ in three dimensions; we discuss its possible experimental implications.
\end{abstract} 
\maketitle
The magnetothermoelectric Nernst-Ettingshausen effect
\cite{ziman2001electrons} has enjoyed renewed interest in recent years, first as a probe of superconducting fluctuations, and more generally, as a novel transport characterization of electronic correlations.
Following the initial work on the cuprates\cite{xu2000vortex} strong magnetothermoelectricity was found in a variety of interesting materials. 
While precise theoretical treatment is lacking for most of these cases, phenomenological descriptions in terms of conventional weak-field quasiparticle transport theory\cite{oganesyan2004nernst,hackl2009nernst,zhang2008anomalous,tewari2009effects} or effective classical hydrodynamic models \cite{ussishkin2002gaussian,podolsky2007nernst} have been used with varied degree of success\cite{behnia2009nernst}. 

In this letter we break from these earlier studies to treat finite field response directly, with no recourse to a low-field regime. Our chief accomplishment is the \emph{exact} expression of the off-diagonal Peltier conductivity, $\alpha_{xy}$, in terms of entropy of free fermionic carriers (see Eqs. \ref{alphagen}, \ref{alphaclassical}, \ref{alpha2d} and \ref{alpha3d}). In two dimensions we prove the universality of this expression (which also applies to Dirac fermions) in the presence of quenched disorder and compare it against available experimental data. 
In three dimensions we obtain, essentially in a closed form, the entire intricate singularity structure in $\alpha_{xy}$ (as a function of magnetic field and temperature) inherited from the Landau level spectrum which bears strong resemblance to thermoelectric phenomenology of graphite\cite{zhu2009nernst}. We also examine the weak field limit, $B\to0$, where we predict a simple, $\alpha_{xy}=-s/B$, dependence on magnetic field and entropy density $s$. 
Quite generally, $\alpha_{xy}$ is somewhat of a less studied and, hence, poorly understood quantity, at least as compared to electrical conductivity or entropy. Thus, our basic result directly linking $\alpha_{xy}$ and entropy (importantly, without invoking the so called ``entropy currents'' used elsewhere  in the literature\cite{cooper1997thermoelectric,zhang2008anomalous}) is useful both for simplifying computations but also for developing intuition.
Even if only approximate in more realistic models (e.g. with inelastic processes ignored by us here), it gives some credence to empirical associations of strong Nernst signatures with singular rearrangements of electronic structure, e.g. phase transitions.

Current flow in the presence of weak electric field and a small thermal gradient is determined by 
\ba
{\bf J} &=& {\bf \sigma} \cdot {\bf E} - {\bf \alpha} \cdot {\bf \nabla} T
\; ,\\
{\bf J^Q} &=& T {\bf\alpha} \cdot {\bf E} - {\bf \kappa} \cdot {\nabla T}
\; ,
\ea
where ${\bf J},{\bf J^Q}, T, {\bf E}, {\bf\sigma}, {\bf\alpha},{\bf\kappa}$ are respectively the charge and energy currents; (slightly modulated in space) temperature and electrical field strength; electrical, Peltier and heat conductivity tensors, respectively. Peltier conductivity is usually extracted from electrical conductivity and thermopower, ${\bf S}={\bf\sigma}^{-1}\cdot{\bf\alpha}$, measured in a zero-current configuration. 

For convenience we consider a particular, Hall ``brick'', sample shape (see Fig. 1) of finite extent along $x-$ and $z-$axis, although our results will be independent of this assumption.
\begin{figure}
\includegraphics[width=2.5in,angle=-90]{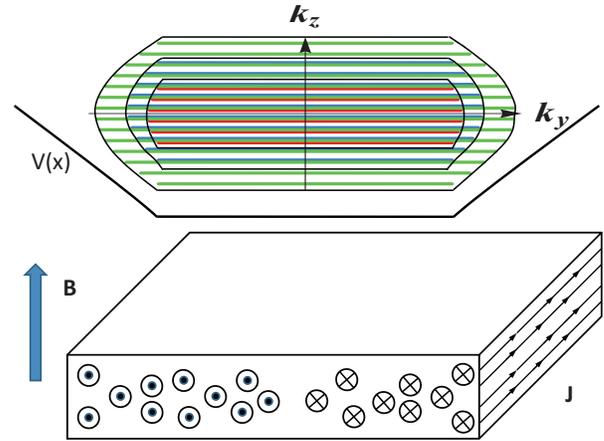}
\caption{(color online) 
Hall ``brick'' in a confining potential $V(x)$ (above) and its Fermi surfaces showing three occupied Landau bands (depicted using different colors). The spectrum is discrete along $k_z$ axis and continuous along $x$. The chiral surface states occupy non-flat portions of the Fermi surface and flow along brick's sides.}
\label{fig:FS}
\end{figure}
Ignoring spin and possible valley quantum numbers for the time being 
the free electron Hamiltonian in a Landau gauge is
\be
{\mathcal H} =
- \frac{\hbar^2}{2m} \left[
(\partial_y - x/\ell_B^2)^2 + (\partial_x)^2
\right]
+ \frac{\hbar^2 k_z^2}{2m} 
+ V(x)
\; ,
\ee
where $m_z$, $m$, $V(x)$, $B$, $-e$ and $\ell_B=\sqrt{\frac{\hbar}{e B}}$ are the electron masses in the $\hat{z}$ direction and $x-y$ plane, confining potential along the x-axis, magnetic field along the z-axis, electron charge and magnetic length, respectively.
The confining potential along the z-axis (not shown explicitly) is assumed to be featureless (hard walls), whereas one along the x-axis, $V(x)$, has a more gradual rise as would be the case in graphite nano-ribbons\cite{sun2008ballistic,novoselov2006unconventional}, for example. Consequently, the wave functions are standing waves along the $z-$axis, labeled with a discrete set of $k_z$'s.
If $V(x)=0$ the spectrum of this problem consists of Landau levels dispersing with $k_z$, $\epsilon_{n,k_z}=\hbar \omega_n+\frac{\hbar^2 k_z^2}{2m_z},$ where $\omega_n=\frac{eB}{m}(n+\nu_0)$, with $\nu_0=1/2$ (but can be different generally\cite{smrcka2009phase}). 
Provided $V(x)$ varies smoothly on scales set by the magnetic length, the spectrum of the Hall brick can be reconstructed by an adiabatic shift of Landau bands at each $k_z$ and $n$, e.g. by absorbing $V(x)$ into a spatially varying $\nu_0$. Thus, the Fermi sea of the Hall brick is a locus of points in the $k_z - k_y$ plane where $\epsilon_{n,k_z}<\mu$, the chemical potential (see Fig. 1).
It is bounded by a set of closed Fermi surfaces each made up of two flat segments in the bulk (with ${\bf k}_{F,n}=\pm(\hat{z}/\hbar)\sqrt{\frac{\mu-\hbar \omega_n}{2m_z}}$) and chiral surface states propagating in $\pm\hat{y}$ directions on opposite sides of the brick 
. While detailed structure of these surface states will depend on surface properties, we generally expect 
them to be robustly present (due to the necessary chirality) and posess interesting transport anisotropies between
directions along the equilibrium current direction ($\pm\hat{y}$) and 
transverse to it, by analogy with chiral metals in layered quantum Hall systems\cite{BalentsFisher:prl96,Chalker:prl95}. The analogy is not precise, since our chiral states are two dimensional, completely delocalized along the surface and smoothly connected to the gapless bulk (layered Hall states typically have mobility gaps in the bulk and strongly one dimensional surface states). 
Their wavefunctions are peaked nearest to the surface for the smallest $k_z\approx \pi/L_z$ component of the Fermi momentum (see fig. 1), which may allow for their characterization via coherent surface tunnelling.
Detailed discussion of these chiral surface states will be presented elsewhere\cite{bergman2009unpublished}.

Bulk transport properties of the Hall brick can be computed easily by applying quantum Hall edge formalism for each transverse mode $k_z$, e.g. following seminal works of Halperin\cite{halperin1982quantized} and Girvin and Jonson\cite{girvin1982inversion} on the Hall bar 
(which itself is a very thin Hall brick with only a single $k_z=\pi/L_z$ state occupied).
We apply both electric field and thermal gradient along the $\hat{x}$ which induces 
net currents 
along the $\hat{y}$ direction (see Eqs. 1 and 2). 
Recall\cite{halperin1982quantized,girvin1982inversion}, that although microscopic current distributions depend sensitively on the details of the confining potential bulk transport coefficients computed by integrating over these distributions are independent of edge specifics, as they must be.  
The three non-zero Hall electric/Peltier/thermal conductivities can be written as $\sigma_{xy}=-\frac{e^2}{h}C_0$, $\alpha_{xy}=\frac{k_B e}{h}C_1$ and $\kappa_{xy}=-\frac{k_B^2 T}{h} C_2$, respectively, with the help of $C_q=\sum_{k_z} c_q(k_z)/L_z$ and 
\be
c_q(k_z) = - 
\sum_n \int_{\hbar \omega_n + \frac{\hbar^2 k_z^2}{2m_z} - \mu}^\infty d \epsilon 
\left( \frac{\epsilon}{k_B T} \right)^q
\frac{ \partial f(\epsilon)} {\partial \epsilon}
\; .
\ee
Here and elsewhere, the Fermi function is denoted by $f(\epsilon) = 1/(1+e^{\epsilon/(k_B T)})$, with $k_B$ the Boltzmann constant.
These coefficients are for electrons, for holes $\sigma_{xy}$ and $\kappa_{xy}$ reverse sign.
Explicit expressions for $c_q$ can be obtained by a further variable change $d \epsilon \to d f$ and a definition 
$f_{n} \equiv f( \hbar\omega_n + \frac{\hbar^2 k_z}{2m_z} -\mu )$, with a familiar result for $c_0(k_z)=\sum_n f_n$ and somewhat less familiar expressions for 
$c_1(k_z)=\sum_n \left[ f_n\log f_n+(1-f_n)\log(1-f_n) \right] $ and 
$c_2(k_z) = \sum_n \left[ \frac{\pi^2}{3}+f_n\log^2(1/f_n-1)-\log^2(1-f_n)-2{\rm Li}_2(1-f_n) \right]$, 
where ${\rm Li}_2(z)$ is the polylogarithm function.
Thus, we find that, up to an overall prefactor, $\alpha_{xy}$ is the entropy per particle added over Landau bands and averaged over transverse modes:
\be
\alpha_{xy}=\frac{k_B e}{h L_z}\sum_n \sum_{k_z}
\left[
f_n \log f_n+ (1-f_n)\log(1-f_n)
\right]
\label{alphagen}
\; .
\ee
Eq. \ref{alphagen} is the basic observation from which the rest of this paper follows (thermal conductivity is left for future work\cite{bergman2009unpublished}).

We start by examining Eq. \ref{alphagen} in the semiclassical regime, with weak scattering and $B\to0$ -- this regime may be realized in some semiconductors\cite{zyryanov1969quantum,fletcher1999magnetothermoelectric}. 
Rewriting Eq. \ref{alphagen} in terms of entropy per volume (i.e. entropy density), $s$, we obtain
\be
\alpha_{xy}=-2\pi \ell_B^2 \frac{e}{h} s=-\frac{s}{B},
\label{alphaclassical}
\ee
akin to $\sigma_{xy}=-\frac{ne}{B}$, where $n$ is the density of particles.
Absent interactions, $\alpha_{xy}$ is even under charge conjugation, while $\sigma_{xy}$ is odd.
At high temperature, for a non-degenerate gas with $s=k_B n$, these results are completely consistent with the purely classical drift in crossed electric and magnetic fields, with entropic force\cite{wu2005giant}, $F=-k_B \nabla T$, and effective electric field is $E=F/e$.
At low temperature, the original $B=0$ Fermi surface of the problem is more or less intact with semiclassical wavepackets of quasiparticles precessing around it. Thus we expect an additional de Haas-van Alphen type magnetooscillations on top of the $1/B$ envelope as well as concomitant Shubnikov-de Hass traces in conductivity. Also, Boltzmann kinetics can be used\cite{oganesyan2004nernst} to interpolate $\alpha_{xy}(B)$ to very low fields where scattering dominates and $\alpha_{xy}\sim B$.

The two dimensional limit of Eq. \ref{alphagen}
is obtained by omitting the sum over $k_z$ modes and setting $f_n=f(\hbar \omega_n - \mu)$
\be\label{alpha2d}
\alpha_{xy}= \frac{e k_B}{h} \sum_{n=0}^\infty 
\left[
f_n\log f_n +(1-f_n)\log (1-f_n)
\right]
\ee
which shows a sequence of thermally broadened nearly symmetric peaks of height $-\frac{k_B e}{h}\log 2$ as a function of $B$ or $\mu$, centered at quantum critical points separating integer Hall plateaux (where $\alpha_{xy}$ shows activated behavior). This structure was discovered and explored previously, albeit in the context of thermpower, by Girvin and Jonson\cite{girvin1982inversion,jonson1984thermoelectric} (although these authors apparently made no identification with the entropy).
We now provide a proof that an exact quantum critical result in Eq. \ref{alpha2d} is, in fact, universal in the presence of disorder. 

For simplicity we consider a setting with generic, short-ranged quenched disorder, strong magnetic field and no boundaries. 
We shall make use of two well established facts about this system. First, strong localization in the presence of magnetic field results in a sequence of integer Hall transitions\cite{huckestein1995scaling}, with 
$\sigma_{xy}(T=0,\mu)=\frac{e^2}{h}\sum_n \theta(\mu-\hbar \Tilde \omega_n)$ 
changing in a discrete series of sharp steps typically located at $\mu=\hbar \Tilde \omega_n$ near the centers of original Landau levels (and $\theta(x<0)=0$ and unity otherwise). Second,  the so-called ``generalized'' Mott formula 
\be\label{generalized_Mott}
C_1(T,\mu) = - \int_{-\infty}^{\infty} d\epsilon \frac{\partial f(\epsilon)}{\partial \epsilon} \frac{\epsilon}{k_B T} C_0(\mu+\epsilon,T=0)
\ee
is an exact integral expression shown previously\cite{smrcka1977transport,jonson1984thermoelectric}  to relate $\sigma_{xy}(T=0)$ and $\alpha_{xy}(T)$.
Substituting the universal behavior of $\sigma_{xy}(T=0)$ into the generalized Mott formula we find Eq. \ref{alpha2d}, with an important adjustment
that $f_n$ should be evaluated at the true critical points, $\mu=\hbar \Tilde \omega_n$ rather than the Landau levels of the problem without disorder. 

Simple as it is, this proof is important on a number of counts. Firstly, it corrects the findings of Jonson and Girvin\cite{jonson1984thermoelectric}, who studied the effects of quenched disorder
using self-consistent Born approximation and concluded that their earlier, clean limit expressions\cite{girvin1982inversion} were not universal. Secondly, universality in the presence of disorder implies that any experimentally observed deviations must be due to experimental uncertainties, electron-electron or electron-phonon interactions, disequilibration effects or other physics omitted thus far. Recent experiments on graphene\cite{checkelsky2009thermopower} found a surprisingly robust confirmation of Eq. \ref{alpha2d} for most plateaux transitions even in the strongly disordered samples, which also hints at the broader universality of Eq. \ref{alpha2d}, e.g. with respect to band structure details. In fact, we have confirmed this universality theoretically for Dirac fermions\cite{bergman2009unpublished}.
Thirdly, the fact that the peaks are centered about the true critical points in the presence of disorder provides a very important additional insight highlighting the difference between the density of states (as measured through magnetization\cite{eisenstein1985density}) and the entropy -- while the former accounts for both localized and extended states, the entropy of localized states is strictly zero, hence activated behavior of $\alpha_{xy}$ is preserved.
Finally, the standard Mott formula 
\be
\alpha(T)=-(\pi^2/3)(k_B^2T/e) A_\alpha\partial\sigma(T,\mu)/\partial \mu,
\ee
with $A_\alpha=1$ is violated in this dissipationless regime, essentially due to singularities of zero-temperature conductivity. 
For concreteness we can use \eqref{alpha2d} and the finite $T$ expression for $\sigma_{xy}$ in the absence of disorder to obtain $A_\alpha(\mu,T)\approx \frac{3}{\pi^2} \beta|\hbar \omega_n-\mu|$ in the activated regime and $A_\alpha=12\log 2/\pi^2\approx 0.84$ precisely at the transition -- observe that Mott formula slightly overestimates $\alpha_{xy}$ near the transition and grossly understimates it away, in the activated regime, as $T\to0$.

Next, we consider the three dimensional regime of large $L_z$, so that a quasi continuum of momenta $k_z$ appears below 
the Fermi level, so we can write simply
\be
\alpha_{xy} = \frac{e k_B}{h} \sum_n \int_{-\infty}^\infty \frac{d k_z}{2\pi} 
\left[ f_n \log f_n + (1-f_n) \log (1-f_n) \right]
\; .
\label{alpha3d}
\ee

One low temperature (Sommerfeld) expansion of this expression can be obtained by linearizing electron dispersion about each of the flat Fermi surfaces with Fermi velocity, $v_{Fn}=
\sqrt{2(\mu-\hbar\omega_n)/m_z}$
\be
\alpha_{xy} = - \frac{e k_B}{h} \frac{\pi^2}{3}  \sum_{n=0}^{n_{\textrm{max}}}\frac{k_B T}{2 \pi \hbar v_{Fn}},
\label{sommer}
\ee
where $n_{\textrm{max}}=[\nu]$, the index of the highest occupied Landau level, is the integer part of  $\nu=\left( \mu/\hbar \omega_c-\nu_0 \right)$.
Defining thermal deBroglie lengths for motion along the $z-$axis, $\lambda_{Tz}=h/\sqrt{2k_B Tm_z}$ and perpendicular to it $\lambda_{T}=h/\sqrt{2k_B T m}$ we can rewrite
\be
\begin{split}
\alpha_{xy} & =
-\frac{e k_B}{h} \frac{\pi^2}{3} \frac{\ell_B}{\lambda_{Tz} \lambda_{T}} 
\sum_{n=0}^{n_{\textrm{max}}} \frac{1}{\sqrt{\nu-n}}
\\ & = 
i \frac{e k_B}{h} \frac{\ell_B}{\lambda_{Tz} \lambda_{T}}
\left[
\xi(1/2,-\nu) -
\xi(1/2,1+n_{\textrm{max}}-\nu)
\right]
\end{split}
\ee
in terms of $\xi$, the Hurwitz zeta functions ($i\equiv\sqrt{-1}$). 
This approximate expression for $\alpha_{xy}$ vanishes linearly with $T$ as $T\to0$, it also diverges at a set of critical fields $B_{Cn}=\frac{\mu m}{e(n+\nu_0)}$ defined by $\hbar \omega_n =\mu$ via one sided root-singularities, due to one dimensional van-Hove singularities appearing every time a Landau band empties, with $k_{Fn}\to0$.
Clearly, Sommerfeld expansion breaks down around these critical points and we need to do better.

\begin{figure}
\includegraphics[width=3.5in]{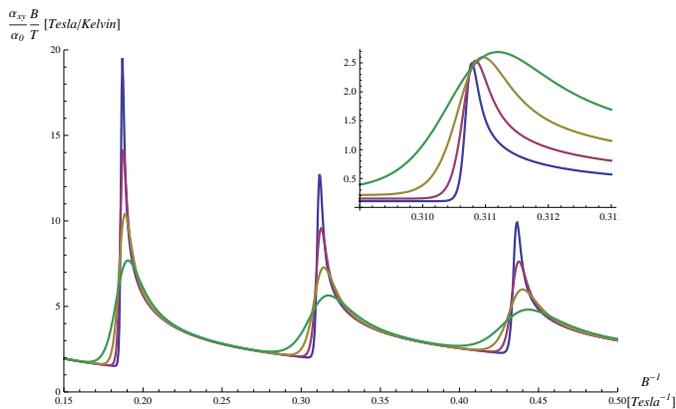}
\caption{Using effective mass parameters $m=0.05m_e$, $m_z=3 m_e$ and carrier density $n=10^{-18}/cm^3$ we plot the expected behavior of $\alpha_{xy}$ across the quantum limit showing both $\sim T$ and quantum critical, $\sim \sqrt{T}$ (in the inset), behaviors. We take $T=0.5,1,2,4 [K]$ for the $\sim T$ plot, and $T=.25,.125,.0625,.03125 [K]$ in the inset (identifyable by the fact that lower temperature gives sharper peaks). We compute $\alpha_{xy}$ in units of 
$\alpha_0 = -\frac{k_B e}{h \lambda}$, where $\lambda = \frac{h}{\sqrt{2 m_z k_B (1 \textrm{Kelvin})}}$.
The quantum critical behavior of the scaling function $F$ is characterized by a temperature and field independent value at the crossing point (signifying the crossing of the Landau band's bottom) and a temperature independent value at the peak $~T$ away from the crossing point. 
}
\label{fig:GNernst}
\end{figure}

To that end we rewrite Eq. \ref{alpha3d}, with $b_n\equiv\beta(\mu-\hbar \omega_n)$ and $f_z\equiv f(z^2-b_n)$,
\ba
\alpha_{xy}&=&-\frac{ek_B}{h\lambda_{Tz}}\sum_n F(b_n),\; \\
F(b)&\equiv&-\int_{-\infty}^{\infty}dz f_z \log f_z + (1-f_z)\log(1-f_z)\\
F\left( b \right) &=& \left\{ \begin{gathered}
\sim 1/\sqrt{b}\; {\rm for}\; b\to\infty \hfill \\
\sim e^{-b}\; {\rm for}\; b\to-\infty \hfill \\
\approx 2\; {\rm at}\; b=0 \hfill \\ 
\approx 2.4\; {\rm for}\; b=b_{min}\approx 1.3\end{gathered} \right.
\ea
in terms of the crossover function $F$ (which is everywhere positive), whose relevant regimes are listed above and displayed in Fig. \ref{fig:GNernst}.
Most importantly, in quantum critical regimes $\hbar e |B-B_{Cn}|/m \lesssim k_B T$ we predict $\sim \sqrt{T}$ variation of $\alpha_{xy}\approx-2(e/h)(k_B/\lambda_{Tz})$.

{\bf Experiments:}
The broad range of materials where our theory may be tested directly includes conventional degenerate and non-degerate semiconductors\cite{fletcher1999magnetothermoelectric}
, semiconductor heterostructures and graphene\cite{wei2009anomalous,zuev2009thermoelectric,checkelsky2009thermopower}, bulk semimetals\cite{behnia2007oscillating,zhu2009nernst}, quantum critical points, esp. ones with pronounced changes in the Hall number\cite{paschen2004hall}. Our discussion here will be limited to some very recent experiments on graphene and graphite -- the relative abundance of data and features make these ideal testbeds.

Several groups have recently examined thermoelectricity in graphene\cite{checkelsky2009thermopower,wei2009anomalous,zuev2009thermoelectric} with data broadly confirming the existence of peaks in thermopower near plateaux transitions. In particular, analysis of the data to extract $\alpha_{xy}$ reveals\cite{checkelsky2009thermopower} a reasonable agreement with Eq. \ref{alpha2d}, although the experimental values appear somewhat smaller than the theoretical prediction in the entire doping range studied.
It is unclear at present whether the observed discrepancy is within experimental uncertainty or if additional theory of inelastic processes (e.g. electron-electron or electron-phonon interactions) is necessary. As the non-interacting theory predicts peaks whose heights are independent of temperature, with widths (in field or gate voltage) narrowing as $T\to0$ it would be of particular interest to measure the temperature dependence of $\alpha_{xy}$.  

Our theory for three dimensions 
appears to capture gross experimental features in graphite\cite{zhu2009nernst} rather well.
The overall scale of the observed signal is reproduced by the effective mass parameters for two dominant electron and hole bands
previously used\cite{du2005metal} in graphite: the estimate of $\alpha_{xy}$ was obtained from thermopower data in Ref. \cite{zhu2009nernst} by taking diagonal resistivity $\rho(B)\approx 2\cdot10^{-5}  B$ $\Omega\cdot m$ per Tesla and neglecting $\alpha_{xx}\rho_{xy}$ contribution.
Most importantly, the experiments clearly demonstrate the existence of asymmetric singularities synchronous with Landau level emptying transitions. 
Well away from these transitions $\alpha_{xy}$ usually vanishes linearly with temperature, as expected by the simple Sommerfeld expansion of Eq. \ref{sommer}.
While showing a clear breakdown of $\sim T$ scaling near the singularities, the existing data\cite{zhu2009nernst} is insufficient to confirm the $\sim\sqrt{T}$ law discussed above, unfortunately.

In conclusion, we identified carrier entropy with $\alpha_{xy}$ exactly, irrespective of temperature. We demonstrated that strong localization in two dimensions enforces this relationship, computed $\alpha_{xy}$ in three dimensions and examined available experimental data in graphene and graphite.
While there is considerable degree of qualitative and quantitative agreement between our theory and the experiments, much work remains to appreciate the full reach and importance of our observation. Clearly, some theoretical treatment of inelastic processes is needed -- while it appears\cite{bergman2009unpublished} that weak electron-electron interactions leave 
our result intact, phonon effects (e.g. drag) on the Nernst thermopower are less understood\cite{zyryanov1969quantum}.
Careful direct measurements (or extractions) of $\alpha_{xy}$ will be of great use as well. Finally, strongly correlated regimes, both at zero\cite{yang2009thermopower} and finite\cite{wang2001onset,cyr2009enhancement} temperatures may also exhibit a version of $\alpha_{xy}$-to-entropy correspondence.

\begin{acknowledgments}
We would like to thank K. Behnia, S. Girvin, L. Glazman and P. Ong for useful discussions. We are especially grateful to K. Behnia for sharing with us his unpublished results and reigniting our interest in Nernst matters.
This work was supported by the National Science Foundation through grant DMR-0803200 and
through the Yale Center for Quantum Information Physics.

\end{acknowledgments}
\bibliography{nernst}
\end{document}